\input{aipcheck}

\documentclass[
    ,final 
  ]
  {aipproc}

\layoutstyle{6x9}

\begin{document}

\title{SIMP: A Near-Infrared Proper Motion Survey}

\classification{97.20.Vs, 95.80.+p}
\keywords      {}

\author{\'Etienne Artigau}{
  address={Gemini Observatory, Southern Operations Center, c/o AURA, Casilla 603, La Serena, Chile}
}
\author{David Lafreni\`ere}{
  address={Department of Astronomy and Astrophysics, University of Toronto, 50 St. George Street, Toronto, ON M5S 3H4, Canada}
}
\author{Ren\'e Doyon}{
  address={D\'epartement de Physique and Observatoire du Mont M\'egantic, Universit\'e de Montr\'eal, C.P. 6128, Succ. Centre-Ville, Montr\'eal, QC, H3C 3J7, Canada}
}
\author{Lo\"ic Albert}{
  address={Canada-France-Hawaii Telescope Corporation, 65-1238 Mamalahoa Highway, Kamuela, HI 96743}
}
\author{Jasmin Robert}{
  address={D\'epartement de Physique and Observatoire du Mont M\'egantic, Universit\'e de Montr\'eal, C.P. 6128, Succ. Centre-Ville, Montr\'eal, QC, H3C 3J7, Canada}
}
\author{Lison Malo}{
 address={D\'epartement de Physique and Observatoire du Mont M\'egantic, Universit\'e de Montr\'eal, C.P. 6128, Succ. Centre-Ville, Montr\'eal, QC, H3C 3J7, Canada}
}

\begin{abstract}
SIMP is a proper motion (PM) survey made with the Observatoire du Mont M\'egantic (OMM) wide-field near-infrared camera CPAPIR at the CTIO 1.5 m and OMM 1.6 m telescopes. The SIMP observations were initiated in early 2005, are still ongoing and, to date, have covered 28\% of the sky at high galactic latitudes. The PMs of the sources detected are determined by comparing their measured positions with those listed in the 2MASS point source catalog, giving a time baseline of 4 to 10 years. The 5 $\sigma$ uncertainty on the relative SIMP and 2MASS astrometry is 1$^{\prime\prime}$, equivalent to a PM lower limit of 0.125-0.250$^{\prime\prime}$/yr, or a tangential velocity limit of $15-30$ km/s at 25 pc. Up to the 2MASS magnitude limit (J$\sim$16.5), T dwarfs are found out to $\sim$25 pc, while L dwarfs may be found as far as 100 pc away.
\end{abstract}

\maketitle
\section{Observations and analysis}
The recent all-sky near-infrared (2MASS, DENIS \citep{Skrutskie2006, Epchtein1997}) and visible (SDSS \citep{York2000}) surveys have enabled the discovery of the majority of known field brown dwarfs (BDs); $\sim$530 L-type and $\sim$146 T-type BDs. Much of our knowledge of the local BD space density, and sub-stellar mass function, comes from these surveys. However, the current census of BDs in the solar neighborhood remains biased because many searches for bright-BD candidates were based on near-infrared color criteria designed to avoid contamination from M dwarfs, which have $JHK_s$ colors similar to those of BDs straddling the L/T transition (L8-T4). 

\begin{figure}[!tbp]
  \includegraphics[height=.23\textheight]{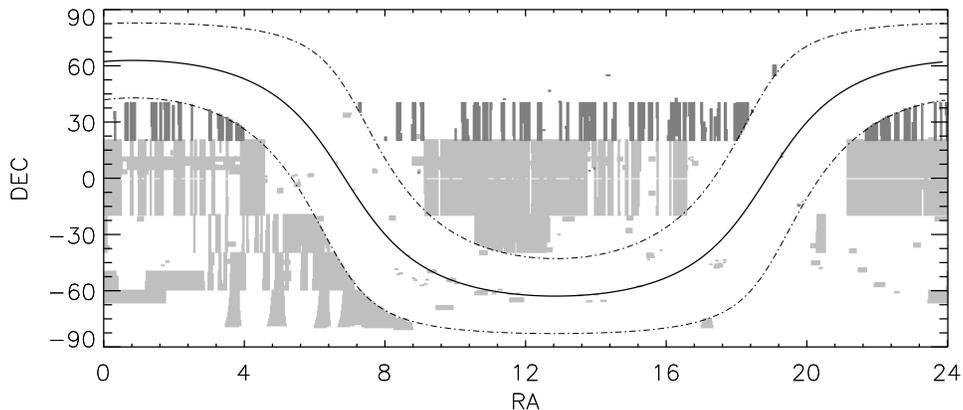}
  \caption{Map of the current sky coverage of SIMP. The survey targets regions of the sky with galactic latitudes greater than 20$^{\circ}$ (dashed line), with additional targeted observations around nearby stars. Regions covered at the CTIO 1.5 m telescope are shown in light grey and regions covered at the OMM 1.6 m telescope in darker grey.}
\label{fig_a}
\end{figure}

Proper motion (PM) is an efficient mean of discriminating BDs from M dwarfs. For a given apparent magnitude, BDs are much closer and therefore have, on average, show larger PMs. We have undertaken a near-infrared PM survey (SIMP: {\it Sondage Infrarouge de Mouvement Propre}) with the wide-field near-infrared camera CPAPIR at the CTIO 1.5m /OMM 1.6m telescopes (FOV of $34.8^{\prime\prime}\times34.8^{\prime\prime}$ at the CTIO and $30^{\prime\prime}\times30^{\prime\prime}$ at the OMM) . We use the 2MASS survey as a first epoch to identify high-PM BD candidates. The integration used was $3\times8$ s exposures for a depth of $J=17.2$ at $5\sigma$, slightly deeper than the 2MASS $J-$band completeness limit. A highly automated pipeline was developed to handle the large amount of data (more than 140 000 raw frames!) generated. So far SIMP has covered $\sim$28\% of the sky at high galactic latitudes (see figure \ref{fig_a}). 

The first steps of the data reduction include flat-fielding of the raw images, bad pixel masking, subtraction of a running sky median, and median combination of the images at each pointing. The resulting images are then undistorted to minimize the residuals between the measured positions of field stars and their positions in the 2MASS point source catalog (PSC). Objects in the images are matched one-to-one with objects in the PSC. We use $I$-band photometry from the SuperCosmos Sky Survey (SSS\citep{Hambly2001}) or, if available, SDSS colors, to further discriminate potentially cool objects. A source is considered a BD candidate if the following criteria are satisfied: PM $>0.1^{\prime\prime}$/yr and detected at $I$ with $I-J$ $>$ $3.5$ (later than M8) or undetected at $I$ with lower limit of $I-J$ $>$ $3$, or $PM$ $>$ $0.2^{\prime\prime}$/yr and undetected at $I$. These criteria are strongly exclusive of M dwarfs. Indeed, late M dwarfs (M5V-M8V; $10.2$ $<$ M$_{J}$ $<$ $11.5$; $2.5$ $<$ $I-J$ $<$ $3.4$) with $J$ magnitudes at the 2MASS completeness limit and typical velocity dispersion of $\sim25$ km/s should be located at $100-170$ pc with PM $<$ $0.05^{\prime\prime}$/yr. Brighter M dwarfs should be detected in $I$ in the SSS and will be rejected from our candidate list.

\section{Followups \& Highlights}
\begin{figure}[!tb]
  \includegraphics[height=.25\textheight]{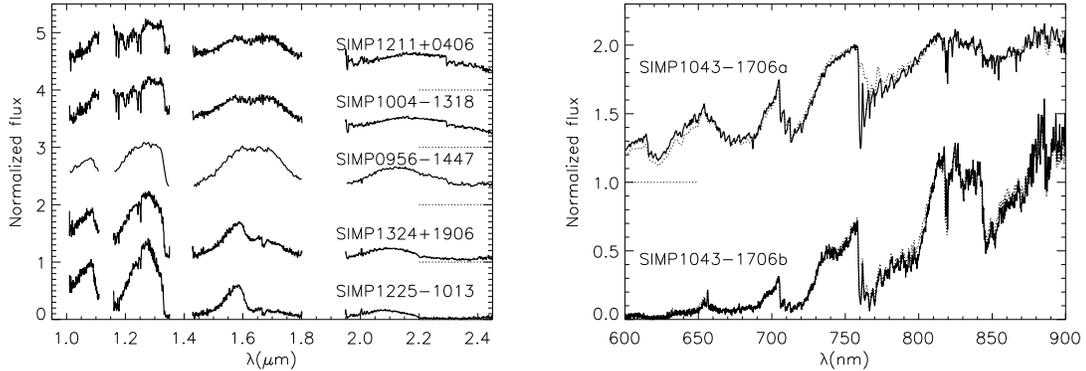}
  \caption{The {\bf left panel} shows the near-infrared spectra of sample of five L and T dwarfs found by the SIMP. The {\bf right panel} shows the $600-900$ nm spectra of both both components of SIMP1043ab. The spectra of both components are overplotted with a comparison template (dotted lines); the M4V spectra from the Pickles \citep{Pickles1998} spectral library (top) and the M7.5 2MASS J0348+2344 from Kirkpatrick et al. \citep{Kirkpatrick1999} (bottom).}
\label{fig_b}
\end{figure}

The SIMP follow-ups have been done along three lines. First, near-infrared low or moderate resolution spectral-typing of candidate L and T dwarfs was done at either the Gemini (using NIRI and GNIRS, $R\sim500$), the IRTF telescope (using SpeX, $R\sim150$, $R\sim250$ and $R\sim750$ depending on the target brightness) or at the OMM using the SIMON near-infrared spectro-imager ($R\sim50$). Among the very first objects uncovered by the SIMP was SIMP0136+09 \citep{Artigau2006}, a bright ($H=12.77$) T2.5 dwarf surpassed in brightness only by $\epsilon$ Indi Bab. Up to now we have spectroscopically confirmed over 15 T dwarfs and more than 60 Ls. Figure \ref{fig_b} shows near-infrared spectra of a sample of five of these objects; see also table \ref{tab_a}.

The large dataset of PMs obtained by SIMP allows to search for BDs that are comoving companions to main sequence stars. These pairs are of particular interest as 1) coeval companions to a main sequence star provide benchmark objects to test brown dwarf evolution models; 2) they are often tight binaries themselves, opening the prospect of a dynamical mass measurements within a few years \citep{Burgasser2005}; and 3) confirming the mass and separation of these systems has a bearing on low-mass star formation models, given their potentially low binding energies. Comoving pairs are followed-up at the Gemini Observatory using both GMOS-N and GMOS-S. Spectroscopy of both the primary and secondary components are obtained over the $600-900$ nm spectral range. This wavelength interval is preferred as it provides important age, gravity and activity indicators such as H$\alpha$, Li I at 670.9 nm and the Na doublet at 818~nm. SIMP uncovered more than $20$ comoving pairs having an M7 to mid-L secondary.

\begin{table}[!tb]

\begin{tabular}{l c ccc}
%\begin{tabular}{lcccc}
\hline
\tablehead{1}{r}{b}{Designation}

  & \tablehead{1}{c}{b}{$J$}
  & \tablehead{1}{c}{b}{$H$}
  & \tablehead{1}{c}{b}{$K_s$}
  & \tablehead{1}{c}{b}{SpT}
   \\
\hline
SIMP095608-144706   & $16.28\pm0.10$ & $15.06\pm0.09$ & $14.22\pm0.06$ & L$9.5$\\
SIMP100440-131818   & $14.69\pm0.04$ & $13.88\pm0.04$ & $13.36\pm0.04$ & L$3.5$\\
SIMP121130+040609   & $15.53\pm0.06$ & $14.55\pm0.04$ & $14.02\pm0.08$ & L$2.5$\\
SIMP122559-101341   & $16.43\pm0.10$ & $16.33\pm0.22$ & undetected     & T$6.5$\\
SIMP132407+190627   & $15.77\pm0.07$ & $15.47\pm0.12$ & $15.41\pm0.16$ & T$4.5$\\\hline

SIMP104325-170606   & $15.65\pm0.06$ & $15.06\pm0.06$ & $14.61\pm0.11$ & M$7.5\pm0.5$\\
SIMP104323-170603   & $11.48\pm0.02$ & $10.89\pm0.02$ & $10.60\pm0.03$ & M$3.5\pm0.5$\\
\hline
\end{tabular}
\caption{A sample of the new objects uncovered by the SIMP. $JHK_s$ photometry from the 2MASS PSC.}
\label{tab_a}

\end{table}

Figure \ref{fig_b} shows the spectra of both components of such a comoving pair, SIMP1043ab, with a $16.90^{\prime\prime}$ separation and spectral types of M$3.5\pm0.5$ and M$7.5\pm0.5$. From the Hawley et al. \citep{Hawley2002} spectral type - M$_{J}$ relation, we find a distance modulus of $3.8\pm0.6$ and $4.6\pm0.2$ for SIMP1043a and SIMP1043b, respectively. The mild inconsistency suggests that either SIMP1043a is an unresolved double or that SIMP1043b is underluminous compared to the bulk of field late Ms. The system is at a distance of 55-75 pc, implying a physical separation of 900-1300 AU. With a total mass of $\sim0.36$ M$_\odot$ (0.26 M${_\odot}$ and 0.098 M${_\odot}$ for components a and b, respectively, using the Baraffe \citep{Baraffe1998} models), this systems lies at the edge of the mass/separation diagram of known binaries (see fig. 6 in  Burgasser et al. \citep{Burgasser2007}).

Finally, $i$-band photometry of all high-PM candidates lacking SDSS and SSS detection is obtained at the CFHT with MegaCam under poor seeing but photometric conditions. To date, 223 candidates have been followed-up with single 200 s exposures.

 Since SIMP BD candidates are inherently bright and local (Ls within $\sim$100 pc and Ts within $\sim$25 pc), many of the new SIMP L and T dwarfs will eventually have their trigonometric parallax measured. Our survey is complementary to surveys like UKIDSS and CFHTLS \citep{Delorme2008b} which will find many faint (i.e. distant) T dwarfs for which detailed spectroscopic study are costly in observing time even on 8-m class telescopes. The nature of this program opens the door to exciting serendipitous discoveries. The CFBDS found the first two ammonia dwarfs in 1000 square degree (CFBDS0059 \citep{Delorme2008} and ULAS0034 \citep{Warren2007} and independently identified in the CFHTLS dataset); at $J=18$ and 18.2, these objects would have been undetected in our survey, but considering that we covered an area more than 10 times larger, we can expect to detect similar but closer objects. At a depth of $J=16.8$, we cover the same volume as a 1000 square degrees survey with a depth of $J=18.3$ and can therefore realistically expect to uncover an ammonia dwarf similar to CFBDS0059 or ULAS0034 within our sample.

\begin{theacknowledgments}
The authors wish to sincerely thank telescope operators, both at CTIO and Mont M\'egantic Observatory, for their indefectible support during the long SIMP nights.
\end{theacknowledgments}

\bibliographystyle{aipproc}

\end{document}